\documentclass[conference]{IEEEtran}
\IEEEoverridecommandlockouts
    \usepackage{comment}                                                       
\usepackage[autostyle]{csquotes}
\IEEEoverridecommandlockouts                              

 \usepackage[pdftex]{graphicx}
\usepackage{booktabs}
\usepackage{array}
\newcolumntype{P}[1]{>{\centering\arraybackslash}p{#1}}
\usepackage{cite}
\usepackage{amsmath,amssymb,amsfonts}
\usepackage{algorithmic}
\usepackage{graphicx}
\usepackage{textcomp}
\usepackage{xcolor}
\usepackage{bbding}

\def\BibTeX{{\rm B\kern-.05em{\sc i\kern-.025em b}\kern-.08em
    T\kern-.1667em\lower.7ex\hbox{E}\kern-.125emX}}
\begin{document}
\title{End-to-End Service Level Agreement Specification for IoT Applications\\
\thanks{The first author is funded by King Saud University, Riyadh, SA}
}
\author{
    \IEEEauthorblockN{Awatif Alqahtani\IEEEauthorrefmark{1}\IEEEauthorrefmark{2}\Envelope, Yinhao Li\IEEEauthorrefmark{1}, Pankesh Patel\IEEEauthorrefmark{3}, Ellis Solaiman \IEEEauthorrefmark{1}\Envelope,  Rajiv Ranjan\IEEEauthorrefmark{1}}
    \IEEEauthorblockA{\IEEEauthorrefmark{1}School of Computing Science, Newcastle University, Newcastle, UK
    \\\{a.alqahtani, y.li119, ellis.solaiman, raj.ranjan\}@ncl.ac.uk}
    \IEEEauthorblockA{\IEEEauthorrefmark{2}Natural and Engineering, College of Applied Studies and Community Service, King Saud University, Riyadh, SA
    }
      \IEEEauthorblockA{\IEEEauthorrefmark{3}Fraunhofer CESE, College Park, Maryland,  USA
    \\ppatel@CESE.fraunhofer.org}
}

\maketitle

\begin{abstract}

The Internet of Things (IoT) promises to help solve a wide range of issues that relate to our wellbeing within application domains that include smart cities, healthcare monitoring, and environmental monitoring. IoT is bringing new wireless sensor use cases by taking advantage of the computing power and flexibility provided by Edge and Cloud Computing. However, the software and hardware resources used within such applications must perform correctly and optimally. Especially in applications where a failure of resources can be critical. Service Level Agreements (SLA) where the performance requirements of such applications are defined, need to be specified in a standard way that reflects the end-to-end nature of IoT application domains, accounting for the Quality of Service (QoS) metrics within every layer including the Edge, Network Gateways, and Cloud. 
In this paper, we propose a conceptual model that captures the key entities of an SLA and their relationships, as a prior step for end-to-end SLA specification and composition. Service level objective (SLO) terms are also considered to express the QoS constraints. Moreover, we propose a new SLA grammar which considers workflow activities and the multi-layered nature of IoT applications. Accordingly, we develop a tool for SLA specification and composition that can be used as a template to generate SLAs in a machine-readable format. We demonstrate the effectiveness of the proposed specification language through a literature survey that includes an SLA language comparison analysis, and via reflecting the user satisfaction results of a usability study.\\
\end{abstract}

\section{Introduction}In IoT environments, devices (e.g., sensors, actuators, and cameras) sense, capture, and send behaviors of the physical world as raw data over computer networks, to the Edge layer and/or the Cloud layer for further processing. Edge and Cloud layers perform computational and analytical operations (e.g., filtering, analyzing, detecting, etc.) on the received data in order to make automatable actions on physical environments and ultimately forward visualized results to end-users. 
The Edge layer typically contains a small-scale datacenter to perform lightweight tasks. In contrast, the Cloud layer consists of large scalable distributed pools of configurable resources, for performing intensive tasks on historical and real time
data, on demand \cite{2}. It allows users to submit jobs for computing, storing, analyzing as well as handling the heterogeneity of data and devices \cite{2}. 
\begin{figure*}[thpb]
      \centering
\includegraphics[width=0.82\linewidth ]{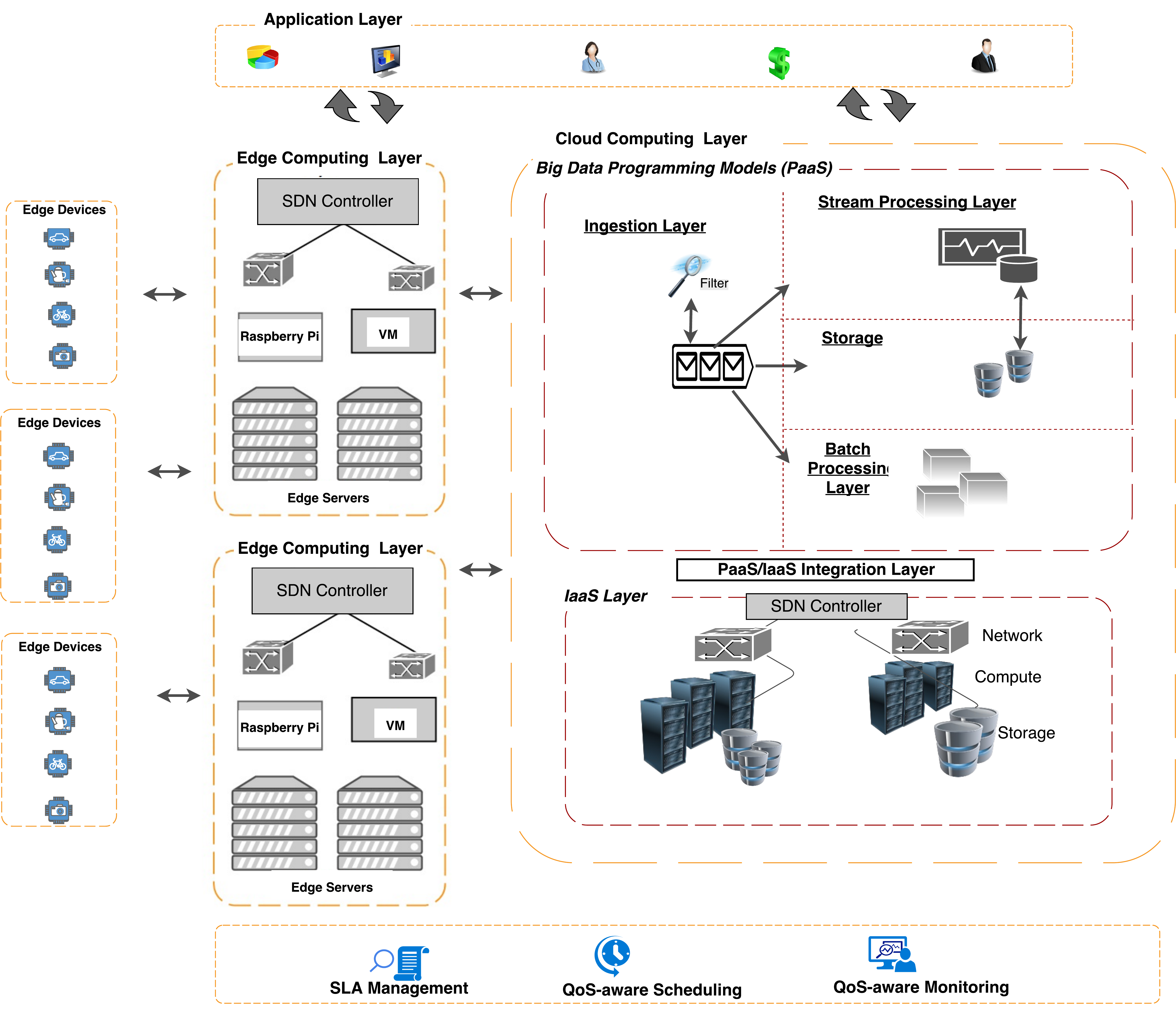}
      \caption{Reference IoT Architecture}
      \label{arc}
   \end{figure*}
\subsection{Motivation and research problem}
Figure 1 illustrates a reference architecture that reflects the multi-layer nature of an IoT application. As can be seen from the figure, one of the main challenges of building IoT applications is their potential complexity. Because of this complexity, integrating distinct technologies and platforms in a way that minimizes failures and guarantees application users a high quality of service (QoS) is not an easy task. 


Many IoT applications are also time sensitive. For example, let us consider a Remote Health Monitoring Service (RHMS) where patient data is collected from different resources (e.g. heartbeat sensors, smart cameras and mobile accelerometers). The filtered data is then transferred to a big data processing platform within the cloud layer for further analysis. 
One of the Service Level Objectives (SLO) of this RHMS could be to \emph{“detect urgent cases within \texttt{Y} time units and notify emergency services within \texttt{X} time units of detection”}. Any unacceptable delay in the transfer of the data for this application might have serious consequences. In order to achieve this SLO, the service provider/s should have in place mechanisms and guarantees on the availability of the service, and on important time constraints by which any critical data must be transferred. At the very least, any hint of performance degradation or failure within the application should be monitored, investigated and tracked to its root cause\cite{31}. Unfortunately, the current generation of application monitoring tools are not capable of this fine grained monitoring required by such IoT applications. 

As a first step toward building IoT applications, one must be able to specify the QoS metrics for an IoT application comprehensively within standardized Service Level Agreements (SLA) that can be understood by all stakeholders involved. 
To emphasize the importance of standardizing SLA for IoT applications, consider a scenario where an IoT application administrator would like to find the best set of providers for the services that matches his/her requirements for developing the desired IoT application. Because IoT applications have a multi-layered architecture, IoT administrators need to consider different categories of providers (e.g., Network provider, Cloud provider, Edge provider) and find the best candidate for each category. In order to be able to communicate requirements with various potential providers, it would be extremely useful to be able to make use of standardized terminology that describes consumer requests as well as provider offers. Such standardization would enable the process of selecting the best candidate services to be automated. For example, the most popular cloud providers (e.g., AWS, MS Azure, Oracle), currently provide take-it-or-leave-it SLAs for their services. When customers need to compare such SLAs from different providers to select the most suitable, they need to do it manually. IoT applications can potentially be much more complex than cloud applications, and therefore such a comparison becomes more difficult. Therefore, standardizing the way SLAs are described for both service consumers as well as service providers, would be an important step towards automating service provider selection. In addition, standardizing machine readable IoT SLAs is also an important step towards automating the process of IoT application deployment, monitoring, and dynamic reconfiguration. Once an IoT application has been deployed, it is important to continuously monitor that the application is adhering to what has been agreed upon and to be able to dynamically reconfigure the application on the fly as needed to ensure that those QoS requirements are met\cite{9}.  

Standardizing the SLA specification for IoT applications is challenging due to a number of factors \cite{7}:
(1) the multi-layered nature of end to end IoT (edge device layer, edge computing layer, cloud layer), (2) several metrics are required to capture the performance of software and hardware components of IoT applications (e.g.,  data freshness at the edge devices and latency of stream processing at the Cloud layer), and (3) dependencies within each of the metrics across IoT layers (e.g., data rate of stream processing at the Cloud layer is affected by sampling rate of the edge devices). 
It is well known that SLA specification languages for various application domains do indeed exist \cite{11}~\cite{12}~\cite{13}~\cite{14}~\cite{16}~\cite{17}. 
However, in their current format, to our knowledge, none of these languages are capable of accommodating the unique characteristics of the IoT domain with its multi-layered nature. In other words, there is an absence of consideration for requirements of all layers (end-to-end) that form an IoT application. 


 

\subsection{Contributions} 
The main aim of this research is to propose a new end-to-end SLA specification language for IoT whilst taking into consideration the challenges presented above. We summarize our contributions as follows:
\begin{itemize}
\item a new conceptual model that captures the knowledge base of IoT specific Service Level Agreements, by expressing the key entities of the IoT ecosystem and the relationships between those entities within SLA context.
\item a new multi-layered grammar to reason about SLA for IoT applications.
 \item a tool for specifying and composing standardized end-to-end SLA constraints with a comprehensive vocabulary, which provides a fine-grained level of specification of user requirements.
 \end{itemize}
 
The remainder of this paper is organized as follows: Section \ref{conceptual} introduces our IoT based SLA conceptual model. The SLA grammar is presented in \ref{grammar}. We demonstrate the tool for specifying and composing end-to-end SLAs for IoT applications and generating machine readable SLAs in section \ref{tool}. Section \ref{relatedwork} reviews related work, and provides a comparison of similar approaches with respect to a number of important criteria. We evaluate our work via a user study in section \ref{evaluation}. Conclusions and future research directions are presented in section \ref{con}.


\section{End-to-End SLA Conceptual Model for IoT Applications}\label{conceptual}
An End-to-end IoT ecosystem, considers all components through which application data is flowing. Components can be hardware, software and/or humans. 
From a Quality of Service (QoS) specification perspective, end-to-end IoT SLAs should consider requirements of entire resources (hardware and software) that are cooperating to deliver the IoT application. This starts from capturing the data, and ends with querying and/or storing the results of any performed analysis, in addition to any other activities, which vary depending on the use case scenario. For example, as depicted in the conceptual model of Figure \ref{uml}, the SLA considers the requirements for all activities that are involved in the use case scenario. To specify the requirements on an end-to-end basis, the conceptual model has considered what services (sometimes referred to within the text as software resources), are required for each activity, and where the services can be deployed. Therefore, the model considers  the infrastructure resources (e.g., IoT devices, Edge resources, and Cloud resource), as well as the services (e.g., sensing service, and real-time analysis service), which can be deployed on the infrastructure resources. In the following section we describe the concepts covered in the conceptual model, and then the relationship between those concepts. 

Our SLA conceptual model for IoT applications is presented in Figure \ref{uml}. In the proposed conceptual model, we refer to the reference architecture (Figure \ref{arc}). The conceptual model is composed  of the following entities: 
\begin{figure*}[thpb]
      \centering
\includegraphics[width=0.80\linewidth]{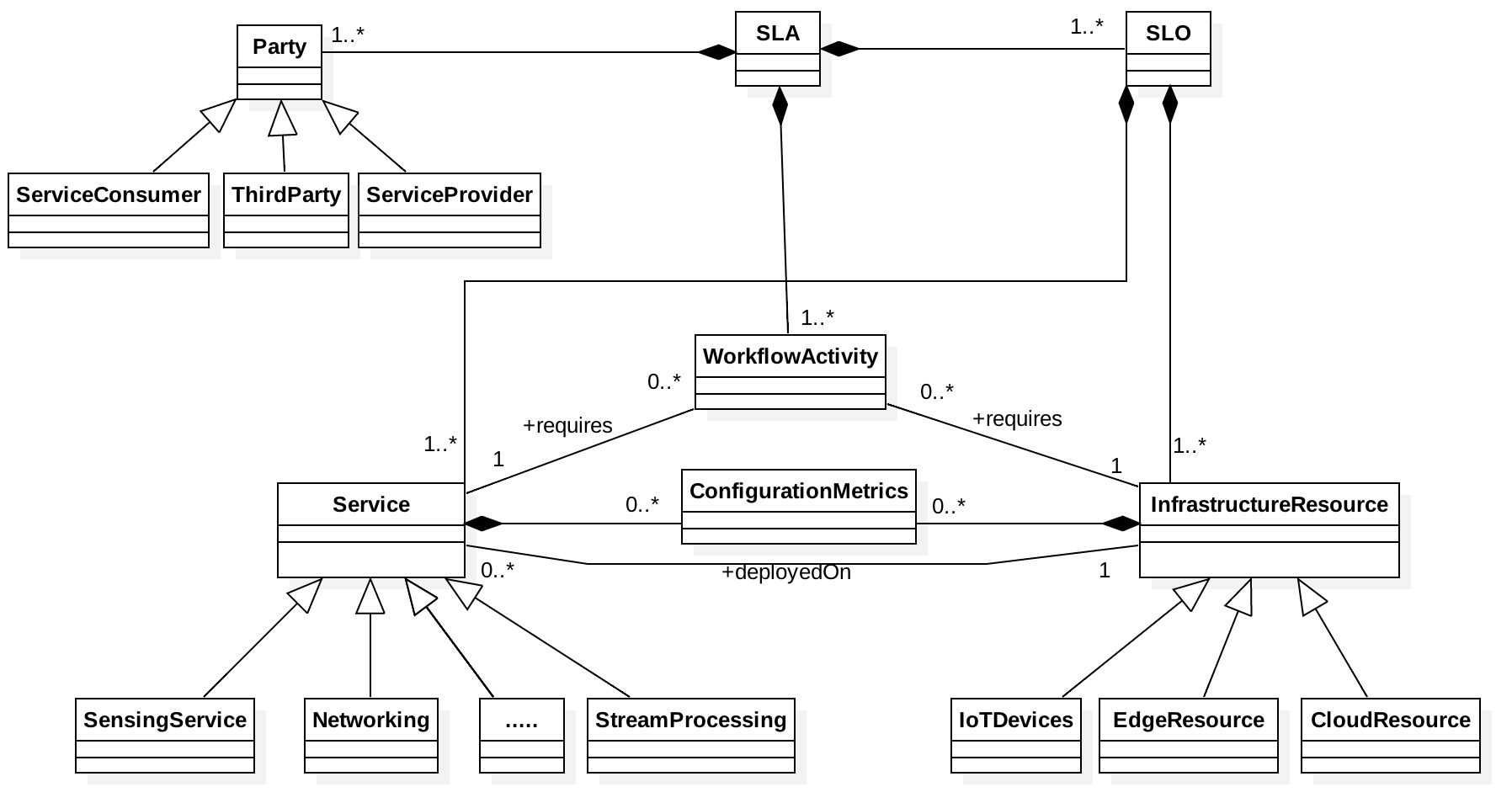}
      \caption{SLA conceptual model  for IoT application which  captures the key entities of an SLA  and their relationships}
      \label{uml}
   \end{figure*}
\begin{enumerate}

\item SLA: includes the basic data such as the title of the SLA, ID of the SLA, type of the application (i.e smart home, smart health, etc.) as well as start and end dates.
\item Party: describes an individual or groups involved in the SLA. It usually contains companies or judicial entities that are named in the SLA \cite{8}. For example, in an RHMS, the parties could be the hospital management, patient, network provider and cloud resource providers.
\item SLO:  provides quantitative means to define the level of service a customer can expect from a provider. It expresses the intention(s) of an agreement for both the application, and any involved services and infrastructure resources. It quantifies the required value of a QoS metric. For example, an SLO (at the application level) of the RHMS scenario,  could be the response to urgent cases within \texttt{Y} unit time.  The QoS metric in this example is \texttt{response time} and the constraint is less than \texttt{Y} unit time. Furthermore, SLO parameter can be used to specify an SLO for lower level services, for example at the data ingestion service, an SLO can be: \emph{"ingest data with latency less than \texttt{Z} unit time"}. For an infrastructure resource such as CPU of a VM, an SLO can be: \emph{"CPU utilization is greater than 80\%"}.  
\item Workflow Activity: IoT applications have certain activities that are required to be considered as part of the application requirements to function correctly. For example, in the RHMS, one of the possible workflow activities is capturing interesting data, analyzing real-time data, and storing interesting results in a database (e.g., SQL or NoSQL).
\item Service: To achieve SLOs at the application level, it is important to establish adequate cooperation between particular services under the SLO constraints. For example, in the RHMS, to detect urgent cases within Y time unit, it is necessary to transfer data from sensors to the ingestion service and to process data on the fly using stream processing services.
Here we list the most common services that can cooperate in order to deliver SLOs of an IoT application. 
\begin{enumerate}
\item  Sensing service: collects data from IoT devices and sends the collected data through a communication protocol to a layer above. It specifies the number of sensors, type of sensors and when to collect the data. For example, in a RHMS, a heartbeat sensor attached to the chest and an accelerometer as a hand-wrist device, reflect the patient’s health state continuously or periodically based on what has been specified within the SLA for this service.
\item  Networking service: communicates the collected data from one layer to another. For example, in the RHMS, home gateway uses the network to deliver collected data to the Cloud for further analysis under certain bandwidth requirements.
\item  Ingestion service: ingests data from many data producers, and then forwards it to subscribed/interested destinations such as storage and analysis services under certain requirements such as throughput limit.
\item  Batch processing service: receives data from resources such as ingestion layers, appends them to the master data set and then computes batch views. For example, in the RHMS, to predict urgent cases it is important to run machine learning algorithms on historical patient records in order to recognize patterns regarding certain health issues and establish a predictive model. The predictive model can be used later with real-time data of current patients in order to detect patients with particular health issues. Batch views can be computed/queried within response time constraints as specified by consumers/subscribers.
\item  Stream Processing Service: processes incoming data from data resources such as an ingestion service to compute real-time views. For example, collected data is processed on the fly, and if the analysis shows an abnormality such as a high heartbeat rate, then appropriate action is required, such as sending an ambulance. However, to observe the greatest value of real-time data, consumers/subscribers can specify certain requirements such as the acceptable delay limit for computing/querying real-time views.
\item  Database service (SQL and NoSQL databases): is used by ingestion, batch and stream processing services to persist or retrieve data. It stores data, batch views and real-time views as intermediate or final data sets. Consumers can specify their requirements on the service such as query response time and specify whether data encryption is required.
\end{enumerate}
\item  Infrastructure resource: provides the required hardware for computation, storage and networking, which are essential to deploy/run the above-mentioned services. The infrastructure resource can be IoT device, Edge resource, Cloud resource.
\begin{enumerate}
\item  IoT device: includes device/object with intelligence ability to actuate on/reflect the physical worlds.
\item  Edge resource: allows for data processing at the edge network. Border routers, set-top boxes, bridges, base stations, wireless access points, edge servers, etc. These are examples of edge resources and these components can be equipped, to support edge computation, within certain capabilities \cite{10}.
\item  Cloud resource: provides  Infrastructure as a Service (IaaS) and, mostly, is located geographically far from the end devices/users \cite{10}.
\end{enumerate}
\end{enumerate}

\begin{figure}[thpb]
      \centering
\includegraphics[width=1\linewidth, scale=0.70]{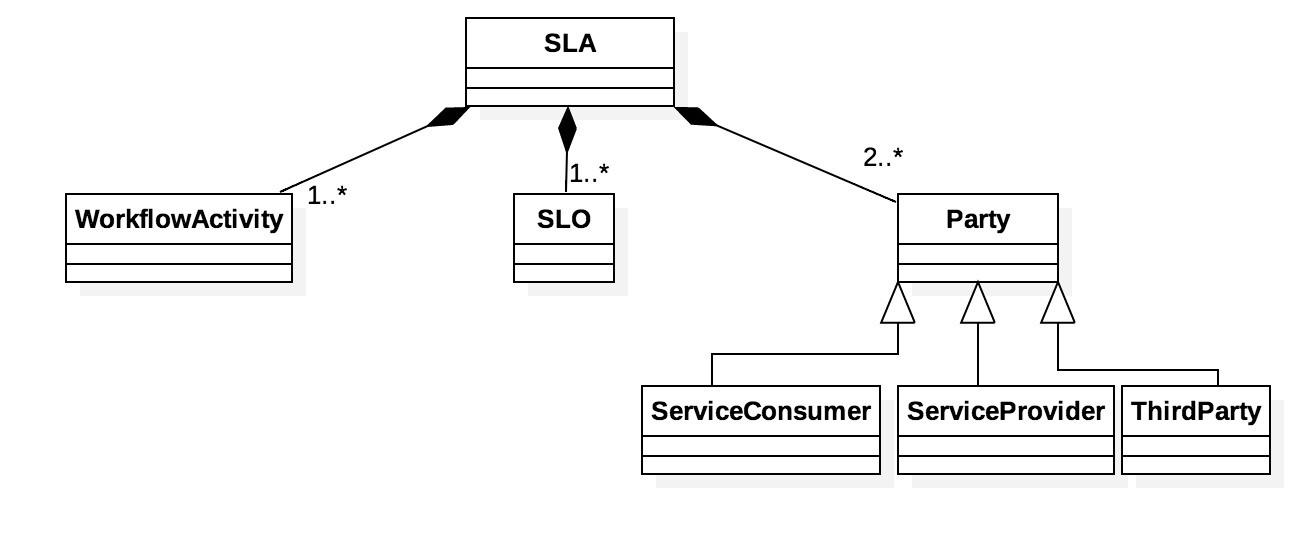}
      \caption{The relationship between SLA, SLO, WorkflowActivity and Party entities}
      \label{composed}
   \end{figure}

The relationships between the above-mentioned entities, which are depicted in the conceptual model (Figure \ref{uml}) are as follows: there is one-to-many relationship between the SLO and the SLA entities to express one or more of the QoS requirements at the application level. Therefore, each SLA entity has a composition relationship with the SLO entity. An example of SLO at application level could be \emph{"end-to-end response time of the application should be less than \texttt{Y} time unit"}.   
Additionally, an IoT application has a set of workflow activities (e.g., capture an Event Of Interest (EoI), analyze real-time data), which cooperate to deliver the application. Therefore, there is a composition relationship between the \texttt{SLA} and \texttt{WorkflowActivity} entities.  Furthermore, an SLA has parties who are responsible for providing, consuming and/or playing third party roles. Figure \ref{composed} depicts the relationship between \texttt{SLA}, \texttt{SLO}, \texttt{WorkflowActivity} and \texttt{Party} entities. 
   
\begin{figure}[thpb]
      \centering
\includegraphics[width=1\linewidth]{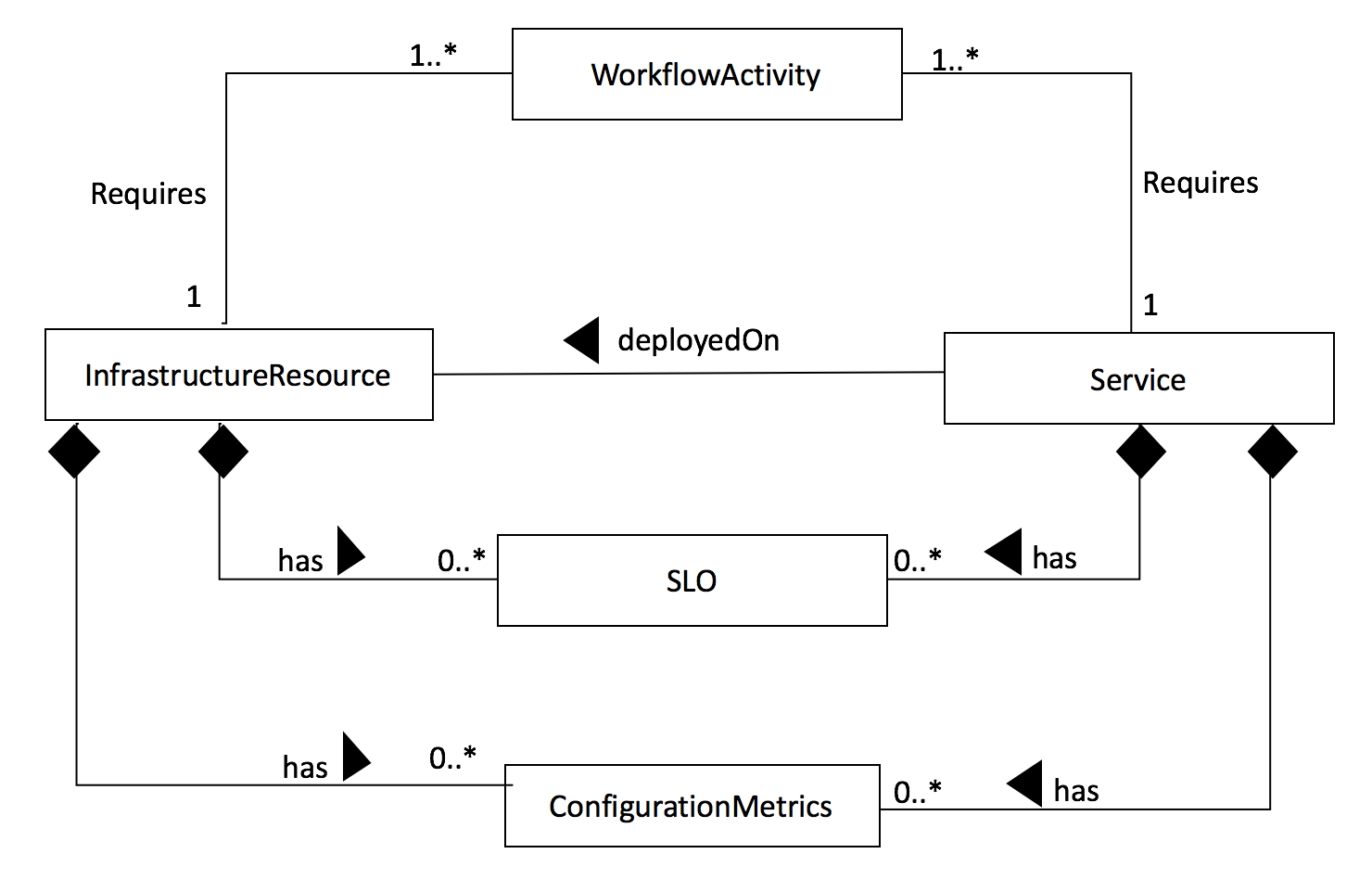}
      \caption{The relationship between  WorkflowActivity, InfrastructurResource, Service, SLO and ConfigurationMetric entities}
      \label{resource}
   \end{figure}

Each workflow activity requires a service (e.g., sensing service, networking service, stream processing service). Each service is deployed on one of the infrastructure resources (for example; edge devices, edge resource, cloud resource). Each one of the services and infrastructure resources has one or more SLOs (e.g., high level of data freshness objective of sensing service and high CPU utilization of VM). Furthermore, each one of the services and infrastructure resources has one or more configuration metrics (e.g., the sample rate of the sensing service and number of CPUs per VM of the cloud resource). Therefore, there is an association relationship between \texttt{InfrastructurResource},  \texttt{Service}, and composition relationship between \texttt{InfrastructurResource},  \texttt{Service}, \texttt{SLO} and \texttt{ConfigurationMetric} entities (Figure \ref{resource}). 

Achieving the SLOs of both services and infrastructure resources has an impact on achieving SLOs at the application level. For example, in the RHMS, an SLO ( SLO$_{app1}$) for urgent cases that require a response within less than \texttt{Y} unit time, is an SLO at the application level which involves many activities such as analyzing real-time data. Analyzing real-time data requires a stream processing service that has an acceptable level of latency, and if the stream processing service exceeds the acceptable level of latency, then SLO$_{app1}$ might be violated. 

\section{SLA Grammar of IoT Applications}\label{grammar}
One of our main objectives is providing a machine-readable SLA specification that can be used by an application orchestrator for automatically deploying IoT applications, and monitoring adherence to the QoS requirements. Table \ref{table1} shows the syntax of our proposed language, which is formally defined in the Extended Backus Naur Form (EBNF): 

\begin{table*}
\centering
\caption{SLA Grammar for IoT Applications }
\includegraphics[width=0.99\linewidth]{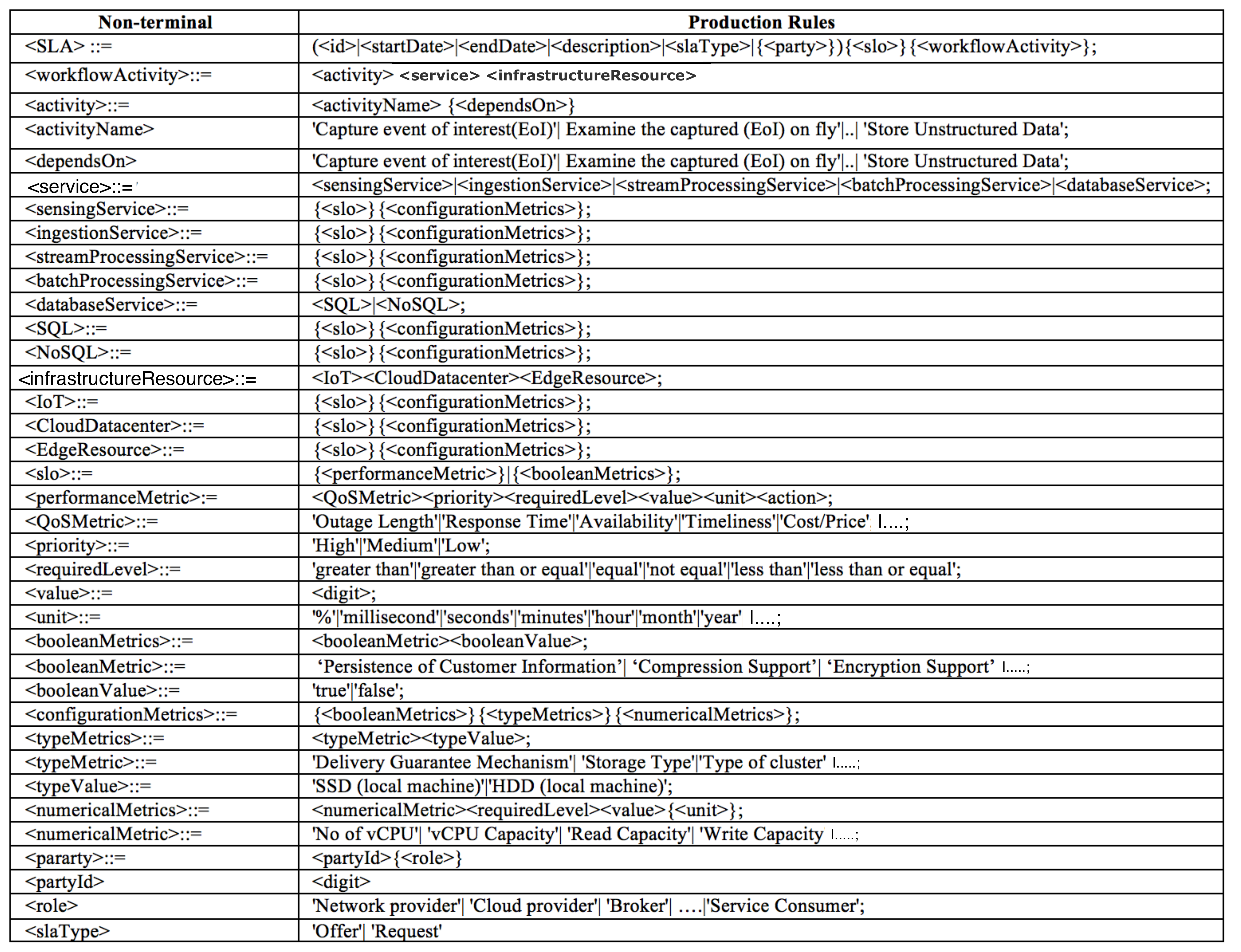}
\label{table1}
\end{table*}

The SLA has the following elements: $<id>$, $<description>$,$<type>$, $<party>$, $<startDate>$, and $<endDate>$, are elements to describe basic information related to the SLA.
Each SLA consists of at least one service level objective $<slo>$ to express the required QoS level at the application level (e.g., in the RHMS, \texttt{response Time} less than 2 minutes). It also contains the priority level (e.g. high, medium, low) of each $<slo>$. For example, in RHMS, response time has higher priority than power consumption, which is not the case with the auto-light building where power consumption has a high priority.
The concept of a $<workflowactivity>$ is used to express the data flow activities of an IoT application (e.g. capture the event of interest, large-scale real-time data analysis, and large-scale historical data analysis).
Each workflow activity is mapped to its required $<Service>$ (such as sensing service, batch processing service) and to $<InfrastructureResource>$ (e.g.,  IoT devices, Edge resources, Cloud resources). Each service and infrastructure resource has its own $<slo>$ and $<configurationMetrics>$. The SLO as mentioned before can express the required level of QoS for each one of the services.
The differentiation between configuration metrics such as $<booleanMetrics>$, $<typeMetrics>$ and/or $<numericalMetrics>$, is based on their values: some metrics have Boolean values, others determine the type of the metric and some have a numerical value. For example,
number of required CPUs, type of clusters are examples of $<numericalMetrics>$, $<booleanMetrics>$ and $<typeMetrics>$, respectively. 

\section{SLA Specification Tool of IoT Application}\label{tool}
We have developed a graphical user interface (GUI)/standalone wizard that can be used by both service consumers and service providers for SLA request and offer creation, respectively. Users of our tool are able to perform the following steps, in sequence (see Figure \ref{sc}):
\begin{figure*}[thpb]
      \centering
\includegraphics[width=0.80\linewidth]{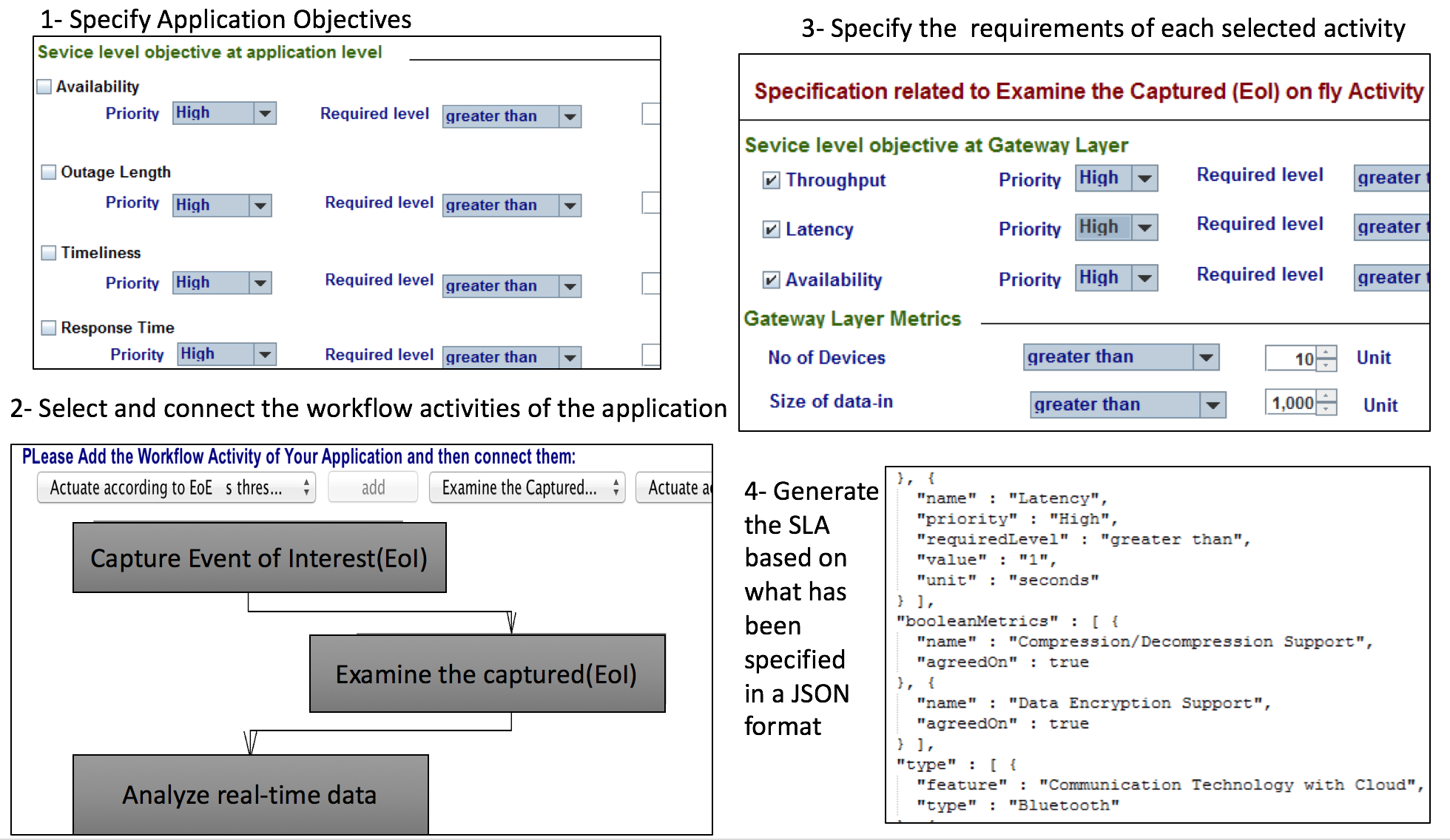}
      \caption{A graphical tool for workflow and SLA specification in IoT applications}
      \label{sc}
   \end{figure*}
 
1) Specify the service level objectives constraints at the application level such as the required/desired level of availability, the time constraints on the response time of the application. For example, to specify the application response time, the user can set the priority (e.g.,  \enquote{high} ), and specify the required level by choosing from a drop-down menu (e.g., \enquote{greater than}) and select the threshold value such as (5) and then select the unit such as \enquote{milliseconds}.

2) Select the workflow activities and connect them to preserve the dependency between the selected activities: user can select the workflow activities based on his/her application scenario. For example if the application is concerned with turning the heater on when the temperature is less than a specific threshold value, then the user can select activities \enquote{Capture EoI}, \enquote{Examine the captured EoI} and \enquote{Actuate based on the captured event’s value} and then connect them to show the sequence of the activities.

3) Map each selected workflow activity to its required service and infrastructure resource: after selecting and connecting the workflow activities, the user can then specify, for each selected activity, the service and the infrastructure resource requirements which host the service. For example, \enquote{Capture EoI} activity requires a sensing service which can be deployed on an IoT device.

4) Specify SLO and configuration requirements for each service and infrastructure resource: After mapping each activity to its required service and infrastructure resource, the user can start specifying the SLO and configuration metrics for both services and infrastructure resources. For example, the user can specify the constraints on \enquote{data freshness} as an SLO requirement of the sensing service as well as specifying \enquote{measurement collection interval} as a configuration metric of the sensing service for \enquote{Capture EoI}.

5) Create SLA document in a JSON format: when users press the \enquote{Finish} button, after specifying their requirements related-to each one of the selected activities, an SLA document will be generated, in JSON format, based on what has been specified.
 
\begin{figure*}[thpb]
      \centering
\includegraphics[width=0.7\linewidth]{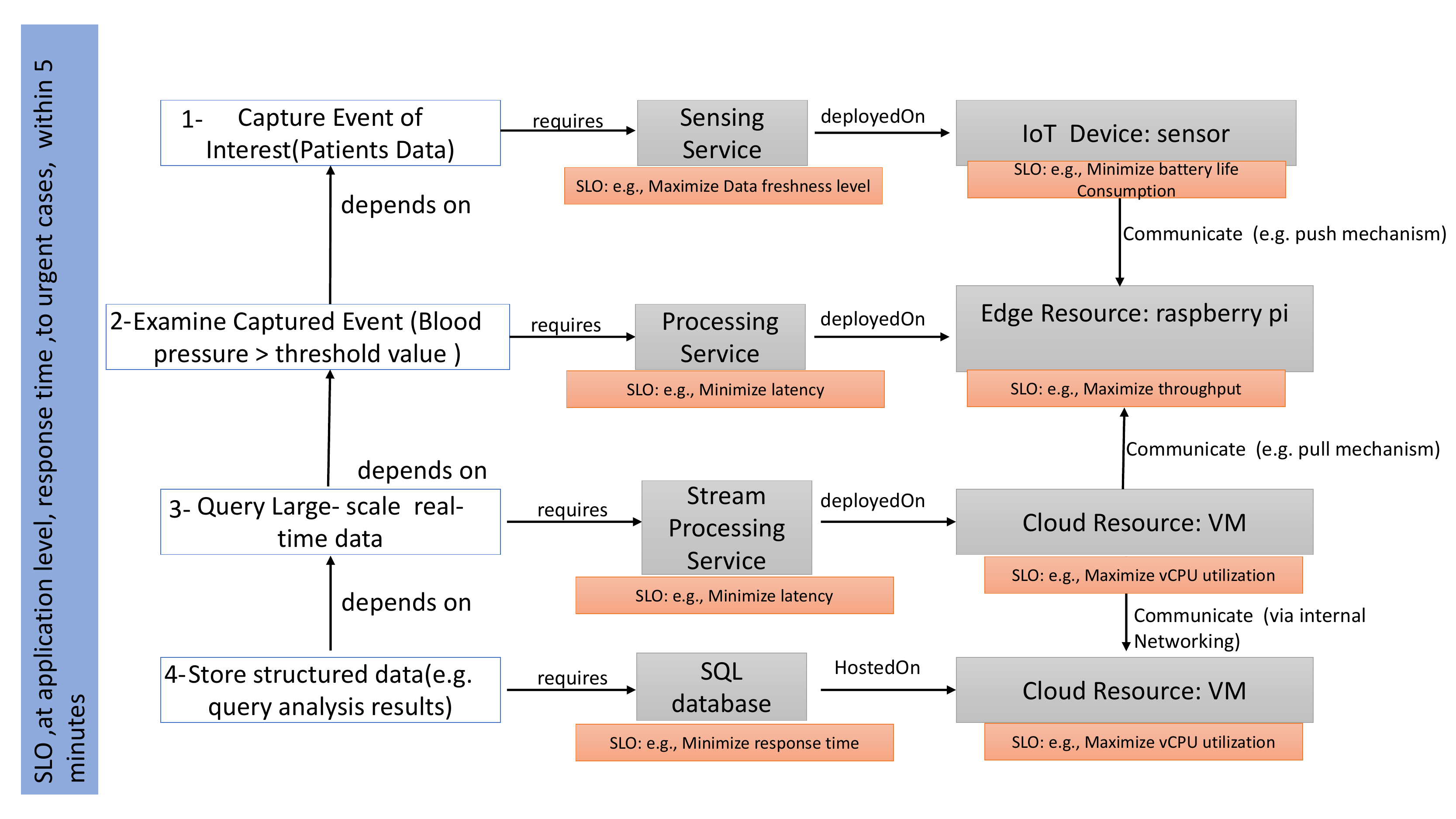}
      \caption{Mapping activities to the required service as well as the infrastructure resource}
      \label{sloexamp}
   \end{figure*}

\begin{figure*}[thpb]
      \centering
\includegraphics[width=0.65\linewidth]{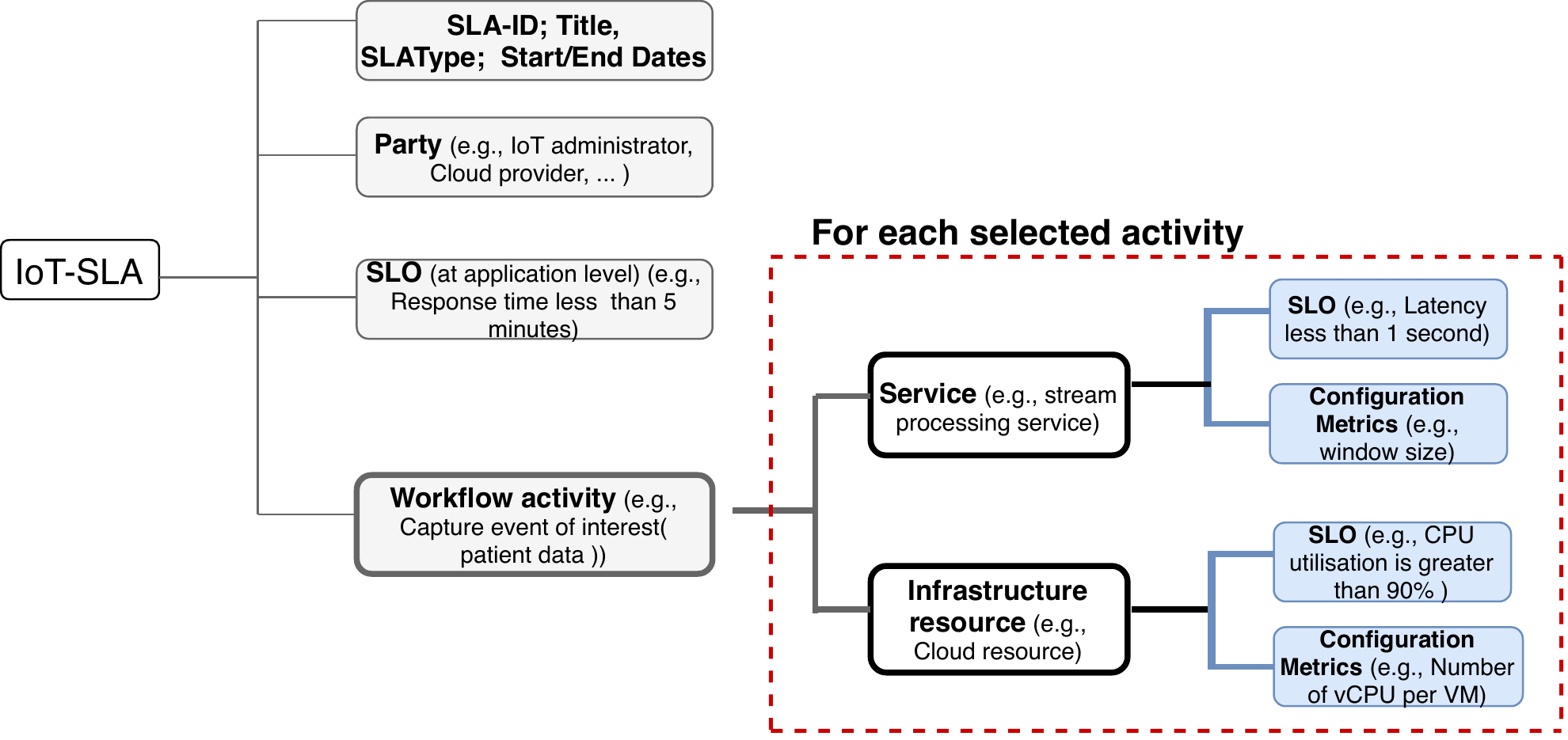}
      \caption{The abstract structure of the main concepts that are considered within the resulted SLA document}
      \label{sloStru}
   \end{figure*}

The tool simplifies and guides the user through the process of generating an end-to-end SLA. It can be used to specify the requirement of different IoT applications. For example, IoT administrator of the RHMS can specify the SLOs of the application such as response time, to urgent cases, should be less than 5 minutes. He/she also will be able to specify the involved activities such as capture EoI (e.g., patients data); examine the captured events (for filtering purpose); analyze real-time data on the fly and store the interesting results. Figure \ref{sloexamp} shows the mapping process for each one of involved activities to the required service as well as the infrastructure resource. It also depicts an example of SLOs related to each one of the required services and the infrastructure resources which are cooperating to deliver the RHMS.
Figure \ref{sloStru} shows the abstract structure of the main concepts that are considered within the resulting SLA document with an example of each concept for clarification purposes.

\section{Related Work}\label{relatedwork}
Studies that are related to our work can be divided into: (a) studies that attempt to identify the most important QoS metrics for one or more of the main layers that are part of the IoT architecture. These are relevant because we aim to consider the key QoS metrics within our SLA language for each of the involved layers within the reference architecture; and (b) Studies that propose SLA languages for various applications, which we compare our own IoT SLA specification language.

\subsection{QoS metrics through the layers}
Within the first Class of studies, QoS metrics for the IoT device layer include the optimum number of active sensors, sensor quality, energy consumption, data volume, trustworthiness, coverage, and mobility \cite{20}\cite{21}\cite{22}\cite{23}. Some of these identified metrics may be inconsiderable for a single edge device\cite{22}, but this is not as trivial as it seems when considering the number of deployed devices that cooperate to deliver a service. For example, a sensor with power consumption equal to 0.9 watts/second seems fine but when a network of hundreds of sensors is deployed then the cumulative value of the power consumption makes a real difference. Network layer QoS metrics such as network availability, network capacity and throughput, mean time to respond, mean time to repair, delay and delay variation, are discussed in \cite{23}\cite{24}\cite{25}\cite{26}\cite{27}. Within the Cloud layer, Throughput and response time are QoS requirements of the data analysis programming models, while CPU utilization, memory utilization, network latency and network bandwidth are examples of QoSs of the infrastructure layer \cite{22}. Research work in \cite{17} identifies a list of metrics related to Cloud computing infrastructures such as CPU, storage size, type of storage (e.g. local storage space), the number of input/output operations on the storage specified for the service, storage bandwidth and operating system. Research in \cite{32}, 
presents an end-to-end performance analysis that identifies key metrics impacting cloud based topic detection and tracking applications. The analysis highlights the complexity of such applications as it captures dependencies between metrics across the cloud layers. At the IoT Application layer, the key requirements vary according to the type/sector that the IoT application is developed for. For example, key requirements for health monitoring applications are robustness, durability, accuracy, precision, reliability, security, privacy, availability and responsiveness \cite{28}\cite{29}\cite{20}. Low-latency is a key requirement in critical and real-time applications \cite{20}\cite{2}, while network utilization and energy efficiency have a high priority in less critical applications such as building automation \cite{20}\cite{30}.
In our work, we considered the most common QoS metrics for all layers of the IoT reference architecture. 

\subsection{Comparison with proposed SLA languages}
In the second class of studies, several projects focus on the development of SLA specification languages \cite{11} \cite{12} \cite{13} \cite{14} \cite{16} \cite{17}. SLA* \cite{11}, CSLA\cite{18}, and SLAC \cite{17} are languages that have been developed for the cloud computing paradigm. The SLA* language is \cite{11} is an abstract syntax for machine-readable SLAs and SLA templates (SLA(T)s). It is a language which is independent of underlying technologies and can be represented by any syntactic format, such as XML or OWL. It provides a specification of SLA(T)s at a fine-grained level of detail. SLA(T) consists of the following sections: an attribute template SLA, the parties to the agreement, service descriptions, variable declarations and the terms of the agreement \cite{11}. Furthermore, the language supports any kind of service and it has been tested on different domains such as enterprise IT and live-media streaming. Nevertheless, specific vocabulary must be defined for each domain \cite{12}. Moreover, the model does not support multi-party agreements \cite{13}.
Authors in \cite{18} provide a Cloud Service Level Agreement (CSLA) to define an SLA for the cloud domain. The CSLA language consists of three sections: the validation period of the agreement; the parties of the agreement; and a reference to the template used to create the agreement. The template defines the service, constraints, the related guarantees, the billing plan and the termination conditions \cite{12}. The concept of fuzziness and confidence is one of the language novelties that considers the dynamic nature of cloud computing. However, there is no formalism for the SLA specification in CSLA \cite{12}. In \cite{17} the authors propose SLAC, which is a language to define a tailored SLA for the cloud domain. The authors specify the syntax of the language as well as the semantics to check the conformance of SLAs. However, SLAC only supports IaaS.
Authors in \cite{16} presented a framework that enables application developers to specify the SLA metric, how it can be calculated, the evaluation period, and constraints to avoid SLA violations using their SLA grammar, named XCLang. However, their main focus is the cloud database.
Due to the limited research efforts that are related to SLA specification language specifically for IoT, we compare our proposed language against the most commonly available service contract languages of Cloud and web services, mentioned above. We use the following main criteria, and present our results in Table 4:

\begin{itemize}
\item IoT Domain: This criterion defines whether a language has been developed for the IoT domain.

\item Ease of use: This criterion can be viewed from the perspective of developers and service consumers. From the service consumer perspective, ease of use is achieved if the user is not required to have much knowledge about how to create the specification in a machine-readable format. From the developer’s perspective, ease of use is determined by whether or not it is written in a machine-readable format. The ease of use criterion is only \texttt{partially} met if just one of these perspectives has been considered.
 \item Support different type of computational resources: Fully supported when a language considers specification requirements of a range of resources such as IoT devices, Edge resources, and cloud resources, and \texttt{partially} supported when it allows for specifying one category of required resources such as only VMs.
 \item Expressiveness: This criterion can be said to be met when the language contains domain-specific vocabulary 
If it does not provide domain-specific vocabulary, then the expressiveness criterion is \texttt{partially} supported.
 \item Syntax: This is supported when there is a formal definition of the syntax e.g. using BNF.
\end{itemize}
Although many SLA specification languages for various application domains do exist, we believe that in their current format they are unable to accommodate the unique characteristics of the cloud-based IoT domain. As can be seen in our comparison Table 4, none of the compared SLA languages provide support for IoT applications.
We have aimed in our specification to consider the most common/typical cloud-based IoT application layers, including data sources, the most common data analysis programming models and computational resources (e.g. IoT, edge resources, cloud datacenters). Furthermore, there are different application models that have different stacks of essential interdependent services. For example, some applications require a certain type of data analysis programming models such as applying data ingestion and stream processing to monitor a patient’s health remotely. On the other hand, other applications that are interested, for example, in computing statistics of a particular vehicle for a month-long period, require ingestion, stream processing, and batch processing data analysis programming models. Therefore, our SLA logic follows the workflow of IoT- based applications, to simplify the process for users (e.g. IoT administrators) to specify their requirements. It enables users to select the workflow of activities for their IoT-based applications as well as to specify their requirements for each service and computational/storage resources (e.g. specify the latency limit of the stream processing service and number of VMs). We have developed a GUI-based tool to enable consumers to specify their requirements. The tool then creates the SLA in a JSON format. By providing a GUI, we ensure the correctness of the SLA specification syntax. Most previous works provide the SLA template in XML format without the support of a GUI, which makes the process of creating a detailed and accurate SLA difficult.

\begin{table*}[!t]
\centering
\caption{Comparison of the SLA languages. Black square() represents a feature supported in the language, empty square represents a partially supported feature and - means not covered.}
\label{tableComp}

\begin{tabular}{p{0.15\linewidth} | p{0.08\linewidth} |p{0.1\linewidth} | p{0.05\linewidth}|p{0.06\linewidth} | p{0.07\linewidth}|p{0.06\linewidth} | p{0.06\linewidth}| p{0.08\linewidth}}
\hline
\textbf{Comparison Features} & \textbf{WASLA} & \textbf{WS-Agreement} & \textbf{SLA*} & \textbf{SLAng} & \textbf{XCLang} & \textbf{CSLA} & \textbf{SLAC} & \textbf{SLA-IoT} \\
\hline

\textbf{IoT Domain} & - &-&-&- &-&-&-&\includegraphics[width=2mm,scale=0.9]{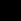} \\
\hline

\textbf{Syntax} & \includegraphics[width=2mm]{black.png} & \includegraphics[width=2mm]{black.png}& \includegraphics[width=2mm]{black.png} & \includegraphics[width=2mm]{black.png} & \includegraphics[width=2mm]{black.png} & \includegraphics[width=2mm]{black.png}& \includegraphics[width=2mm]{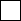} & \includegraphics[width=2mm]{black.png} \\
\hline

\textbf{Expressiveness} & \includegraphics[width=2mm]{empty.png} & \includegraphics[width=2mm]{empty.png}& \includegraphics[width=2mm]{empty.png} & \includegraphics[width=2mm]{black.png} & \includegraphics[width=2mm]{black.png} & \includegraphics[width=2mm]{black.png}& \includegraphics[width=2mm]{black.png} & \includegraphics[width=2mm]{black.png} \\
\hline

\textbf{Ease of use} & \includegraphics[width=2mm]{empty.png} & \includegraphics[width=2mm]{empty.png}& \includegraphics[width=2mm]{empty.png} & \includegraphics[width=2mm]{empty.png} & \includegraphics[width=2mm]{empty.png} & \includegraphics[width=2mm]{empty.png}& \includegraphics[width=2mm]{empty.png} & \includegraphics[width=2mm]{black.png} \\
\hline

\textbf{Support different type of computational resources} & \includegraphics[width=2mm]{empty.png} & \includegraphics[width=2mm]{empty.png}& \includegraphics[width=2mm]{empty.png} & \includegraphics[width=2mm]{empty.png} & \includegraphics[width=2mm]{empty.png} & \includegraphics[width=2mm]{empty.png}& \includegraphics[width=2mm]{empty.png} & \includegraphics[width=2mm]{black.png} \\
\hline

\end{tabular}
\end{table*}

\section{Evaluation}\label{evaluation}
\begin{figure}[thpb]
      \centering
\includegraphics[width=0.95\linewidth]{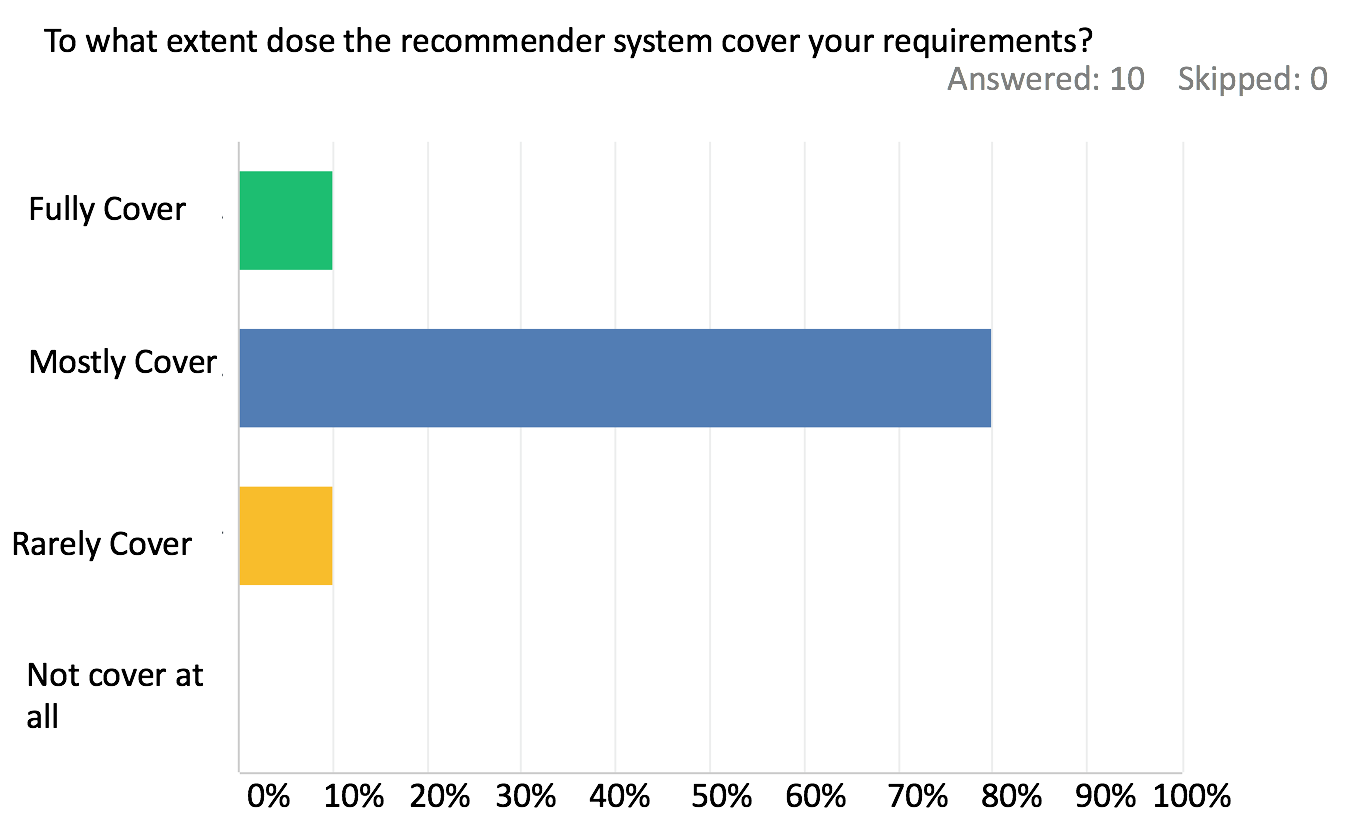}
      \caption{Users' responses regarding to what extent the recommender system covers their requirements.}
      \label{ev}
   \end{figure}

We have developed a context-aware rule-based recommender system; IoT-CANE (Context-Aware recommendatioN systEm). IoT-CANE has been integrated with our specification tool described in section \ref{tool}, and it facilitates incremental knowledge acquisition and declarative context driven knowledge recommendation. This rule-based recommendation system is intended to automatically suggest configuration knowledge artifacts to multiple layers required for users during the IoT resource configuration management processes. Recommended suggestions are generated based on a user-specified context. In the processing layer of IoT-CANE, the admin specifies each resource configuration artifact using the SLA specification tool of section \ref{tool} based on user context information, then stores them into a configuration knowledge database.
In order to evaluate user satisfaction of using the recommender from different perspectives, we conducted a user study with domain experts. The entires results of the user study will be released in a future publication. Importantly, one of the perspectives for the research work in this paper is whether the recommender fully captures user requirements. Figure \ref{ev} shows the user satisfaction for ten participants whose research interest lies on IoT, Networking, Cloud and big data. 80\% of the participants classified their requirements as \texttt{mostly covered},  10\% of them considered their requirements as \texttt{fully covered}, and 10\% of them considered them as \texttt{rarely covered}.

\section{Conclusion and Future Work}\label{con}
The development of an automated end-to-end IoT SLA authoring mechanism, which considers the system requirements of software and hardware components, their related constraints, their interdependencies, and that is machine readable, plays a significant role in automating the deployment, monitoring, and dynamic reconfiguration of IoT applications. The probability of SLA compliance with no need for human intervention is increased by providing an SLA-aware monitoring system. We believe that a machine-readable SLA can be used as a roadmap for system architects and developers. Defining \enquote{SLA offers} and \enquote{SLA requests} using standard vocabularies eases the process of comparing available options and selecting the most suitable SLA offer based on consumer requirements. To this end, we have proposed an SLA specification that reflects the workflow activities of an IoT application and their related requirements in an unambiguous way. Our approach to specifying and composing the end-to-end SLA for IoT applications consists of three main phases. First, a conceptual model represents a knowledge base of SLA specification and composition by capturing the key entities of an SLA and their relationships. Second, a syntax grammar for end-to-end SLAs is derived from the proposed conceptual model. Third, a tool provides a GUI that allows the user to specify SLAs based on the workflow activity of an IoT application, which produces the SLA in a JSON format.  

As part of our future work, we are aiming to represent the knowledge-base of our conceptual model as an ontology. Furthermore, we will develop an SLA-based broker system for IoT applications. The aim of the SLA-based broker system is to receive the generated machine-readable SLA (SLA offers and SLA requests), and find the best candidate that matches user requirements as a step for automating service provider selection. We are also in the initial stages of investigating the development of an IoT monitoring platform that makes use of novel Blockchain and Smart Contract technology~\cite{33}~\cite{34}.



\bibliographystyle{IEEEtran}
\bibliography{ref}

\begin{thebibliography}{10}
\providecommand{\url}[1]{#1}
\csname url@samestyle\endcsname
\providecommand{\newblock}{\relax}
\providecommand{\bibinfo}[2]{#2}
\providecommand{\BIBentrySTDinterwordspacing}{\spaceskip=0pt\relax}
\providecommand{\BIBentryALTinterwordstretchfactor}{4}
\providecommand{\BIBentryALTinterwordspacing}{\spaceskip=\fontdimen2\font plus
\BIBentryALTinterwordstretchfactor\fontdimen3\font minus
  \fontdimen4\font\relax}
\providecommand{\BIBforeignlanguage}[2]{{%
\expandafter\ifx\csname l@#1\endcsname\relax
\typeout{** WARNING: IEEEtran.bst: No hyphenation pattern has been}%
\typeout{** loaded for the language `#1'. Using the pattern for}%
\typeout{** the default language instead.}%
\else
\language=\csname l@#1\endcsname
\fi
#2}}
\providecommand{\BIBdecl}{\relax}
\BIBdecl

\bibitem{2}
M.~DÃ­az, C.~MartÃ­n, and B.~Rubio, ``State-of-the-art, challenges, and
  open issues in the integration of internet of things and cloud computing,''
  \emph{Journal of Network and Computer Applications}, vol.~67, 2016.

\bibitem{31}
E.~Solaiman, R.~Ranjan, P.~Jayaraman, and K.~Mitra, ``Monitoring internet of
  things application ecosystems for failure,'' \emph{IT Professional, IEEE},
  2016.

\bibitem{9}
R.~Ranjan, S.~Garg, A.~R. Khoskbar, E.~Solaiman, P.~James, and
  D.~Georgakopoulos, ``Orchestrating bigdata analysis workflows,'' \emph{IEEE
  Cloud Computing}, vol.~4, no.~3, pp. 20--28, 2017.

\bibitem{7}
A.~Alqahtani, E.~Solaiman, R.~Buyya, and R.~Ranjan, ``End-to-end qos
  specification and monitoring in the internet of things,'' \emph{IEEE
  Technical Committee on Cybernetics for Cyber-Physical Systems}, 2016.

\bibitem{11}
K.~T. Kearney, F.~Torelli, and C.~Kotsokalis, ``Sla*: An abstract syntax for
  service level agreements,'' in \emph{2010 11th IEEE/ACM International
  Conference on Grid Computing}, Oct 2010, pp. 217--224.

\bibitem{12}
R.~B. Uriarte, ``Supporting autonomic management of clouds:
  Service-level-agreement, cloud monitoring and similarity learning,'' Ph.D.
  dissertation, IMT Institute for Advanced Studies Lucca, 2015.

\bibitem{13}
A.~Maarouf, A.~Marzouk, and A.~Haqiq, ``A review of sla specification languages
  in the cloud computing,'' in \emph{2015 10th International Conference on
  Intelligent Systems: Theories and Applications (SITA)}, 2015.

\bibitem{14}
A.~Andrieux, K.~Czajkowski, A.~Dan, K.~Keahey, H.~Ludwig, T.~Nakata, J.~Pruyne,
  J.~Rofrano, S.~Tuecke, and M.~Xu, ``Web services agreement specification
  (ws-agreement),'' in \emph{Open grid forum}, vol. 128, no.~1, 2007.

\bibitem{16}
D.~Stamatakis and O.~Papaemmanouil, ``Sla-driven workload management for cloud
  databases,'' in \emph{2014 IEEE 30th International Conference on Data
  Engineering Workshops}, March 2014.

\bibitem{17}
R.~B. Uriarte, F.~Tiezzi, and R.~D. Nicola, ``Slac: A formal
  service-level-agreement language for cloud computing,'' in \emph{Proceedings
  of the 2014 IEEE/ACM 7th International Conference on Utility and Cloud
  Computing}, ser. UCC '14.\hskip 1em plus 0.5em minus 0.4em\relax IEEE, 2014.

\bibitem{8}
G.~Gaillard, D.~Barthel, F.~Theoleyre, and F.~Valois, ``Sla specification for
  iot operation-the wsn-sla framework,'' Ph.D. dissertation, INRIA, 2014.

\bibitem{10}
R.~Mahmud and R.~Buyya, ``Fog computing: {A} taxonomy, survey and future
  directions,'' \emph{CoRR}, vol. abs/1611.05539, 2016.

\bibitem{20}
C.-L. Fok, C.~Julien, G.-C. Roman, and C.~Lu, ``Challenges of satisfying
  multiple stakeholders: Quality of service in the internet of things,'' in
  \emph{Proceedings of the 2Nd Workshop on Software Engineering for Sensor
  Network Applications}, ser. SESENA '11, 2011.

\bibitem{21}
J.-P. Calbimonte, M.~Riahi, N.~Kefalakis, J.~Soldatos, and A.~Zaslavsky,
  ``Utility metrics specifications. openiot deliverable d422,'' Tech. Rep.,
  2014.

\bibitem{22}
P.~P. Jayaraman, K.~Mitra, S.~Saguna, T.~Shah, D.~Georgakopoulos, and
  R.~Ranjan, ``Orchestrating quality of service in the cloud of things
  ecosystem,'' in \emph{2015 IEEE International Symposium on Nanoelectronic and
  Information Systems}, 2015.

\bibitem{23}
R.~Duan, X.~Chen, and T.~Xing, ``A qos architecture for iot,'' in \emph{2011
  International Conference on Internet of Things and 4th International
  Conference on Cyber, Physical and Social Computing}, Oct 2011.

\bibitem{24}
B.~Li and J.~Yu, ``Research and application on the smart home based on
  component technologies and internet of things,'' \emph{Procedia Engineering},
  2011, cEIS 2011.

\bibitem{25}
\BIBentryALTinterwordspacing
``Practical guide to cloud service agreements version 2.0,'' 2015, [Online;
  accessed 31-March-2018]. [Online]. Available:
  \url{http://www.cloud-council.org/deliverables/CSCC-Practical-
  Guide-to-Cloud-Service-Agreements.pdf}
\BIBentrySTDinterwordspacing

\bibitem{26}
E.~C. Kim, J.~G. Song, and C.~S. Hong, ``An integrated cnm architecture for
  multi-layer networks with simple sla monitoring and reporting mechanism,'' in
  \emph{Network Operations and Management Symposium, 2000. NOMS 2000. 2000
  IEEE/IFIP}, 2000.

\bibitem{27}
B.~Bhuyan, H.~K.~D. Sarma, N.~Sarma, A.~Kar, and R.~Mall, ``Quality of service
  (qos) provisions in wireless sensor networks and related challenges,''
  \emph{Wireless Sensor Network}, vol.~2, no.~11, p. 861, 2010.

\bibitem{32}
M.~Wang, P.~Jayaraman, E.~Solaiman, L.~Chen, Z.~Li, S.~Jun, D.~Georgakopoulos,
  and R.~Ranjan, ``A multi-layered performance analysis for cloud-based topic
  detection and tracking in big data applications,'' \emph{Future Generation
  Computer Systems}, 2018.

\bibitem{28}
O.~Vermesan, P.~Friess, P.~Guillemin, R.~Giaffreda, H.~Grindvoll,
  M.~Eisenhauer, M.~Serrano, K.~Moessner, M.~Spirito, L.-C. Blystad
  \emph{et~al.}, ``Internet of things beyond the hype: Research, innovation and
  deployment,'' \emph{IERC Cluster SRIA}, 2015.

\bibitem{29}
R.~M. Savola, P.~Savolainen, A.~Evesti, H.~Abie, and M.~Sihvonen, ``Risk-driven
  security metrics development for an e-health iot application,'' in \emph{2015
  Information Security for South Africa (ISSA)}, 2015.

\bibitem{30}
M.~V. Moreno, B.~Ãbeda, A.~F. Skarmeta, and M.~A. Zamora, ``How can we
  tackle energy efficiency in iot basedsmart buildings?'' \emph{Sensors}, 2014.

\bibitem{18}
Y.~Kouki, F.~A. d.~Oliveira, S.~Dupont, and T.~Ledoux, ``A language support for
  cloud elasticity management,'' in \emph{2014 14th IEEE/ACM International
  Symposium on Cluster, Cloud and Grid Computing}, 2014.

\bibitem{33}
\BIBentryALTinterwordspacing
C.~Molina-Jimenez, E.~Solaiman, I.~Sfyrakis, I.~Ng, and J.~Crowcroft, ``On and
  off-blockchain enforcement of smart contracts,'' \emph{CoRR}, 2018. [Online].
  Available: \url{eprint arXiv:1805.00626}
\BIBentrySTDinterwordspacing

\bibitem{34}
C.~Molina-Jimenez, I.~Sfyrakis, E.~Solaiman, I.~Ng, M.~W. Wong, A.~Chun, and
  J.~Crowcroft, ``Implementation of smart contracts using hybrid architectures
  with on-and off-blockchain components,'' \emph{arXiv:1808.00093 [cs.SE]},
  2018.

\end{thebibliography}

\end{document}